\documentclass{article}

\usepackage{spconf,amsmath,amsfonts,graphicx}
\usepackage{amssymb,textcomp,mathtools}
\usepackage{bm,upgreek,algorithm,hyperref}
\usepackage{multirow,booktabs,hhline,array}
\usepackage{cite,url,makecell,setspace, xcolor}
\usepackage{enumitem}
\pdfoutput=1

\renewcommand{\vec}[1]{\bm{\mathrm{#1}}}

\title{Multi-Channel Speech Denoising for Machine Ears}
\name{\begin{tabular}{c}
\it Cong Han$^{1,2*\thanks{*~Work done during an internship at X. Current email: ch3212@columbia.edu}}$, E. Merve Kaya$^1$, Kyle Hoefer$^1$, Malcolm~Slaney$^{1,3}$, Simon~Carlile$^1$  \\
\end{tabular}
}
\address{
    $^1$ X, Mountain View, CA, USA \\
    $^2$ Department of Electrical Engineering, Columbia University, NY, USA \\
    $^3$ Google Research, Mountain View, CA, USA
}

\begin{document}
\ninept
\maketitle

\begin{abstract}
This work describes a speech denoising system for machine ears that aims to improve speech intelligibility and the overall listening experience in noisy environments. We recorded approximately 100 hours of audio data with reverberation and moderate environmental noise using a pair of microphone arrays placed around each of the two ears and then mixed sound recordings to simulate adverse acoustic scenes. Then, we trained a multi-channel speech denoising network (MCSDN) on the mixture of recordings. To improve the training, we employ an unsupervised method, complex angular central Gaussian mixture model (cACGMM), to acquire cleaner speech from noisy recordings to serve as the learning target. We propose a MCSDN-Beamforming-MCSDN framework in the inference stage. The results of the subjective evaluation show that the cACGMM improves the training data, resulting in better noise reduction and user preference, and the entire system improves the intelligibility and listening experience in noisy situations.

\end{abstract}

\section{Introduction}
\label{sec:introduction}
Removing noise from speech signals is a good way to improve a user's experiences in noisy environments. New hardware allows for multiple microphones near the ear and the processing power to learn from these signals, which can deliver a better auditory experience. This work describes a system for denoising speech signals captured by a pair of microphone arrays near the ears under noisy conditions. We capitalize on deep neural network (DNN) architectures for speech enhancement, along with multi-channel beamforming.

DNN training requires a large quantity of realistic training data. For speech enhancement, the labeled data is a pair of noisy and clean speech signals. One can create arbitrary amounts of noisy data by adding reverberation and noise; however there is still a mismatch between simulated noisy mixtures and real-world audio due to the complex acoustics of real environments. This could degrade the DNN's performance when it is applied to real-world data.

In this work we instead measured a large quantity of speech and noise signals in a real room, and create mixtures from these recordings. This is good for realism but also includes room tone. An important part of this work is a method for preprocessing the recorded sound to remove this background noise so that it can be used as ground truth.  We demonstrate improvements in noise reduction and listening preference due to the preprocessing.

The rest of the paper is organized as follows. We review related work in Section~\ref{sec:related} and introduce our method in Section~\ref{sec:method}. We describe the experiment configurations in Section~\ref{sec:config}, analyze the results in Section~\ref{sec:result}, and conclude the paper in Section~\ref{sec:conclusion}.

\begin{figure*}[!t]
    \centering
    \includegraphics[width=2.0\columnwidth]{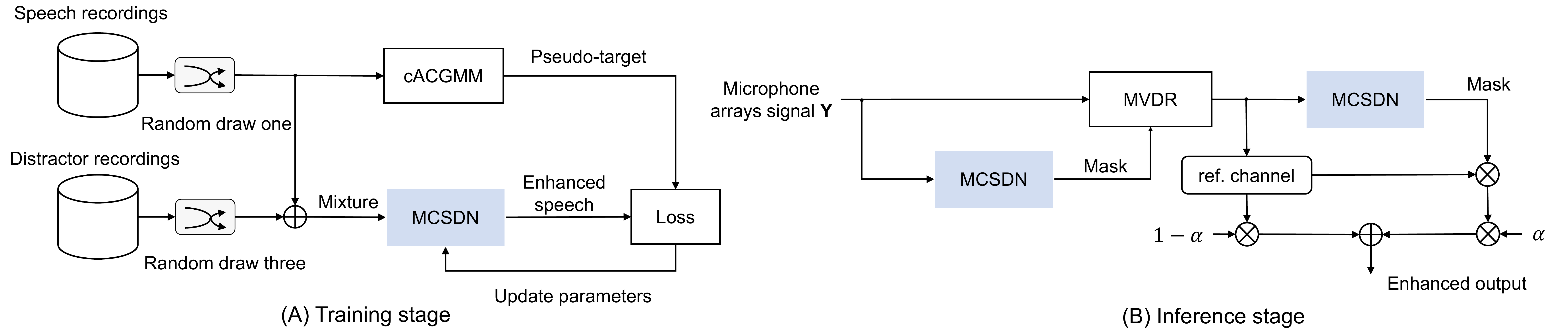}
    \caption{Overview of the speech denoising system. (A) The training stage, and (B) the inference stage. Blue blocks denote the same multi-channel speech denoising network. MVDR is a minimum variance distortionless response beamformer.}
    \label{fig:arch}
\end{figure*}

\section{Related work}
\label{sec:related}
Speech enhancement has been actively studied for decades \cite{boll1979suppression, ephraim1984speech, cohen2001speech,gannot2001signal}. In recent years, deep neural networks have greatly advanced speech enhancement using both supervised \cite{xu2013experimental,weninger2015speech,tan2018convolutional} and unsupervised methods \cite{seetharaman2019bootstrapping, drude2019unsupervised, togami2020unsupervised,nakagome2020mentoring, DBLP:conf/nips/WisdomTEWWH20}. Supervised methods have achieved overwhelming performance, but they require access to ground-truth signals, and thus they can degrade on real recordings which are mismatched with the simulated data used for training. Unsupervised methods overcome these problems by requiring only the noisy speech signals. 

A general category of unsupervised approaches utilizes spatial information to cluster sound sources in space \cite{seetharaman2019bootstrapping, drude2019unsupervised, togami2020unsupervised,nakagome2020mentoring}. The posterior cluster labels can be used as masks to isolate the target speech. An approach using the complex angular-central Gaussian mixture model (cACGMM) \cite{drude2019unsupervised} clusters the signals, and the resulting labels are used as pseudo-target to train a deep clustering model \cite{hershey2016deep}. 

This paper employs cACGMM to extract cleaner speech signals from the recordings to serve as the training target. Our motivation is that we can easily collect moderately noisy recordings without access to ground-truth signals in real scenarios, which can be well processed by the unsupervised clustering methods. Then, we mix several recordings into a much noisier mixture and take advantage of supervised learning to predict the clean speech signals from the mixture. The difference compared to prior work is that we do not apply the clustering model to the noisy mixture directly, because the clustering-based methods perform poorly in challenging conditions where spatial features are smeared by room reverberance and strong background noise, especially diffuse noise with no distinct directional features.


Using a DNN to predict the masks that estimate the spatial covariance, which steers the beamformer toward the target signal, is a popular method to combine DNNs and conventional beamforming methods \cite{erdogan2016improved,nakatani2017integrating,ochiai2020beam}. The linear beamformers effectively keep speech free of nonlinear distortion, which is essential for good perceptual quality of speech in communication. However, the linear beamformer cannot cancel all interference, especially those close in space to the speech source. To reduce the residual noise, the beamforming output can be filtered by the mask used for beamforming \cite{erdogan2016improved} or can be processed by a new post enhancement neural network \cite{wang2021multi} or even more iterations of neural network and beamforming \cite{wang2021sequential}. However, adding a new neural network increases the size of the system, which is undesired for its deployment on hardware with limited capacities, such as hearing aids. In this paper, we employ the exact same multi-channel DNN to predict masks for both the estimation of beamformer weights and post enhancement.

\section{Method}
\label{sec:method}
Figure~\ref{fig:arch} shows an overview of the proposed method for training and inference. During training we use cACGMM to generate a better target to train a conventional multi-channel speech denoising network (MCSDN). In inference, we first apply the pretrained MCSDN, beamforming, and the same MCSDN sequentially. The final denoised signal is a weighted linear combination of the second MCSDN output and the beamforming result.


\subsection{Complex angular central Gaussian mixture model}
\label{gmm}

\begin{figure}[!b]
    \centering
    \includegraphics[width=1.0\columnwidth]{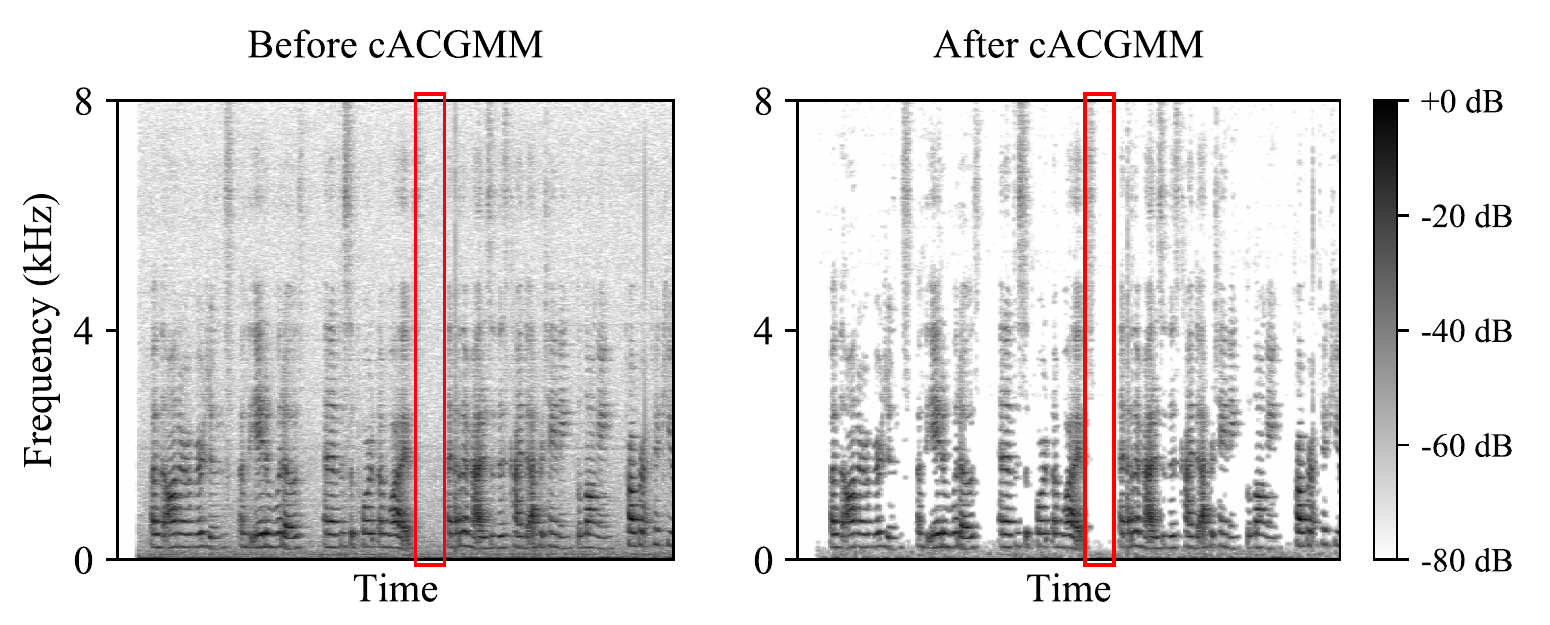}
    \caption{The cACGMM extracts a cleaner speech (right) signal from the recording (left) as the training target. The red rectangle highlights an instance where the background noise is attenuated.}
    \label{fig:p2}
\end{figure} 

Give an M-channel recording $\vec{S} \in \mathbb{R}^{M\times T \times F}$ in the short-time
Fourier transform (STFT) domain, where $T$ and $F$ denote the time frame and frequency bin, respectively, we use the complex angular central Gaussian mixture model (cACGMM) \cite{ito2016complex} to isolate the speech source $\hat{\vec{S} } \in \mathbb{R}^{M\times T \times F}$ and remove noisy sources from unwanted directions. cACGMM models the directional observations $\vec{Z}_{t,f} = \frac{\vec{S}_{t,f}}{||\vec{S}_{t,f}||}$ with a Gaussian mixture model,

\begin{align}
    p(\vec{Z}_{t,f};\Theta_f) = \sum_{k=1}^K{\alpha^{k}_{f}\mathcal{A}(\vec{Z}_{t,f}; \vec{B}_{f}^{k})},
\end{align}
where $\Theta_f = \{\alpha^{k}_{f},\vec{B}_{f}^{k} \forall k \}$ denotes the model parameters, $\{\alpha^{k}_{f}\forall k \}$ is a set of $k$ mixture weights, which are probabilities that sum to 1. $\mathcal{A}(\vec{s}_{t,f}; \vec{B}_{f}^{k})$, a complex angular central Gaussian distribution (cACG) \cite{kent1997data}, models the distribution of $\vec{Z}_{t,f}$ for the component $k$ in the mixture model as follows:

\begin{align}
    \mathcal{A}(\vec{Z}_{t,f}; \vec{B}_{f}^{k}) = \frac{(M-1)!}{2\pi^{M}\text{det}(\vec{B}_{f}^{k})}\frac{1}{(\vec{Z}_{t,f}^H(\vec{B}_{f}^{k})^{-1}\vec{Z}_{t,f})^M}.
\end{align}
We estimate the parameters $\Theta_f$ with the expectation-maximization (EM) algorithm. The posterior probability of $\vec{Z}_{t,f}$ belonging to class k is:
\begin{align}
    \Gamma_{t,f}^k = \frac{\alpha^{k}_{f}\mathcal{A}(\vec{Z}_{t,f}; \vec{B}_{f}^{k})}{\sum_{k=1}^K{\alpha^{k}_{f}\mathcal{A}(\vec{Z}_{t,f}; \vec{B}_{f}^{k})}}.
\end{align}

Since cACGMM models each frequency independently, there can be a frequency permutation problem \cite{sawada2007measuring}, i.e., the same index $k$ in different frequency bins point to different sources. This problem is addressed by permutation alignment \cite{sawada2007measuring}. Finally, we use $\Gamma^s$ as the mask to extract speech, 
\begin{align}
    \hat{\vec{S}} = \vec{S} \odot \Gamma^s,
    \label{eq:cacgmm}
\end{align}
where the superscript s indicates the speech, and $\odot$ denotes element-wise multiplication.


\subsection{MCSDN-Beamforming-MCSDN framework}
\label{dnn1}
We train a multi-channel speech denoising network (MCSDN) based on the temporal convolutional network (TCN) \cite{bai2018empirical} in Conv-TasNet \cite{luo2019conv} to predict a time-frequency mask $\vec{M}^{s} \in \mathbb{R}^{T \times F}$ for the target $\hat{\vec{S}}$ from the multi-channel noisy signal $\vec{Y} \in \mathbb{C}^{M \times T \times F}$. In order to exploit both the spectro-temporal and spatial information, we concatenate the log power spectrogram of the reference channel signal and inter-channel phase differences (IPDs) between the reference channel and other channels as input features, since IPDs indicate which $T$-$F$ bins belong to the same directional source in each frequency band. Specifically, we calculate \text{sin(IPD)} and \text{cos(IPD)} as inter-channel features \cite{wang2018multi, han2020real}. The training objective is defined as:
\begin{align}
    \mathcal L = |\vec{Y} \odot \vec{M}^s - \vec{\hat{S}}|.
\end{align}

We use the estimated mask $\vec{M}^{s}$ from the MCSDN for mask-based beamforming. We employ minimum variance distortionless response (MVDR) beamforming \cite{souden2009optimal}, which is optimized with a constraint that minimizes the power of the noise without distorting the target speech. One solution is:
\begin{align}
    \vec{w}_{f} = \frac{(\vec{\Phi}_{f}^{n})^{-1}\vec{\Phi}_{f}^s}{\text{Trace}({(\vec{\Phi}_{f}^{n})^{-1}\vec{\Phi}_{f}^s})}\vec{u},
    \label{eq:bf_eq}
\end{align}
where $\vec{\Phi}_{f}^{n}$ and $\vec{\Phi}_{f}^{s}$ are the covariance matrices of the speech and noise, respectively:
\begin{align}
    &\vec{\Phi}_{f}^{s} = \frac{1}{\sum_{t}{\vec{M}_{t,f}^s}}\sum_{t}{\vec{M}_{t,f}^s} \vec{Y}_{t,f}\vec{Y}_{t,f}^{H}, \\
   &\vec{\Phi}_{f}^{n} = \frac{1}{\sum_{t}{(1-\vec{M}_{t,f}^s)}}\sum_{t}{(1-\vec{M}_{t,f}^s)} \vec{Y}_{t,f}\vec{Y}_{t,f}^{H},
\end{align}
and $\vec{u}$ is a one-hot vector indicating the reference channel. \textit{H} denotes conjugate transposition. Then, the linear filter $\vec{w}_f \in \mathbb{C}^{M}$ is applied to $\vec{Y}_{t,f} \in \mathbb{C}^{M}$ to generate the beamforming output $\vec{BF}_{t,f}$:
\begin{align}
    \vec{BF}_{t,f} = \vec{w}_f^H \vec{Y}_{t,f}.
\end{align}


Next, we use the trained MCSDN to denoise the beamforming output. To extend the single-channel beamforming output to a multi-channel one, $\vec{BF} \in \mathbb{C}^{M \times T \times F}$, we stack the beamforming output on each channel $[\vec{BF}_{1}, \vec{BF}_{2}, \dots, \vec{BF}_{m}]$ by  shifting the one-hot vector $\vec{u}$ in Equation~\ref{eq:bf_eq} without introducing any new computation. Finally, the MCSDN takes $\vec{BF}$ as input and estimates a speech mask $\hat{\vec{M}}^s \in \mathbb{R}^{T \times F}$ to further denoise the beamforming output:
\begin{align}
    \overline{\vec{S}} = \hat{\vec{M}}^s \odot \vec{BF}.
\end{align}
We can view the pipeline in this way: the first time-frequency mask $\vec{M}^s$ is for mask-based beamforming that results in a less noisy mixture, then the second mask $\hat{\vec{M}}^s$ is to extract the speech from the less noisy signal. However, using the spectral mask estimated by neural networks to extract speech will inevitably cause non-linear speech distortion, which is undesirable for human listeners. We balance the noise reduction and speech distortion by mixing the beamforming output and the neural network output using a gate $\alpha \in [0, 1]$,
\begin{align}
    \widetilde{\vec{S}} = \alpha \cdot \vec{BF} + (1-\alpha) \cdot \overline{\vec{S}}.
\end{align}

\section{Experiment configurations}
\label{sec:config}
\begin{figure*}[!t]
    \centering
    \includegraphics[width=2.0\columnwidth]{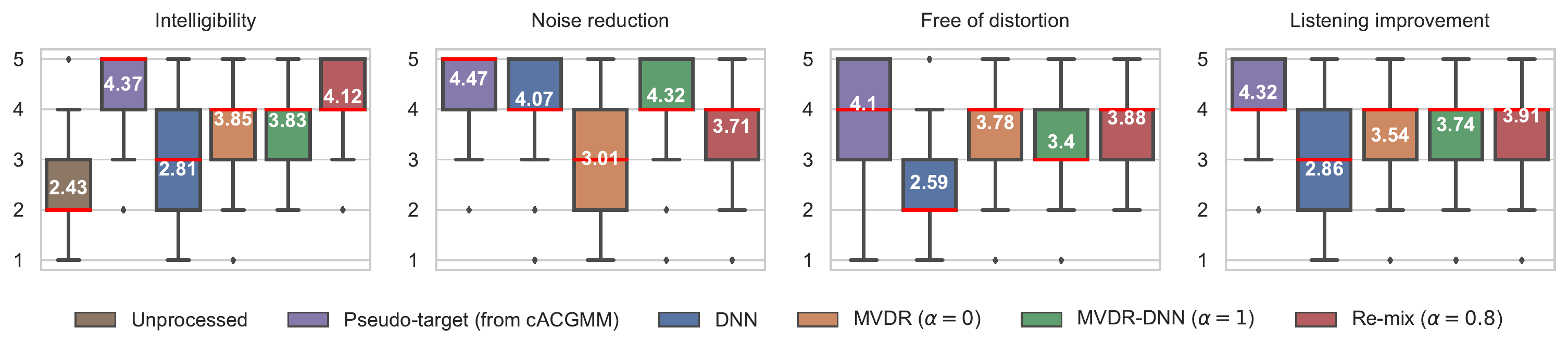}
    \caption{Subjective evaluation results shown as boxplots. Pseudo-target (from cACGMM) is the training target of MCSDN. The red line represents the median (all the ratings are discrete numbers from 1 to 5, so the median is discrete). The number in white denotes the mean. }
    \label{fig:r4}
\end{figure*}

\subsection{Data collection}
\label{dc}
We recorded a collection of in-room speech and ambient sound samples to be used to generate sound mixtures. Each sample captures the real room acoustics, including reverb and background noise. The recording room has the dimensions 7.5 (length) x 3.5 (width) x 3 (height) meters (T60 $\approx 0.37s$). A Bruel and Kjaer Type 4128-C Head and Torso Simulator (HATS) is placed in the center of the room on a motorized turntable, which sits atop a wooden table that spans the majority of the room length-wise. For this experiment, we used two proprietary arrays of 16 microphones, one placed around each of the two ears. Surrounding the HATS are six Genelec 8020D 4" powered studio monitors for audio source playback. All speakers face towards the HATS. The speakers are placed at azimuth values ranging from 0 to 360\textdegree, and the distance from the HATS and elevation from the ground values ranges among 1, 2, or 3 meters. Once placed, the speaker locations are fixed and do not change for the given room.




Playback source data consists of the sound clips from FSD50K \cite{eduardo_fonseca_2020_4060432} ($\sim$50h and 11h from training and test sets) and LibriSpeech \cite{7178964} ($\sim$40h and 11h from training and test sets). We resampled the playback data to 48 kHz, and pre-processsed to trim silence from the beginning and end of each clip. We manually normalized the ound clips so that the clip's db SPL at the speaker is as close as possible to a real-world example for that sound class. The target db SPL values for each sound class have a random variance of $\pm 5$ db SPL. We played each sound clip through a speaker assigned at random, and recorded through the microphone arrays at 48 kHz with 32-bit floating point precision. 

\subsection{Acoustic scene generation}
\label{acg}
For the training and development sets, we generated 12,000 and 4,000 9-second mixtures, respectively. For each mixture, we randomly draw one speech recording and three distractor recordings from the in-room recordings and mix them to simulate challenging environments. We excluded the following broad class labels from FSD50K: speech, alarm, domestic animal sounds, domestic sounds (faucet, cutlery, drawers, etc.). A clip of ambient noise recorded in the room without loudspeakers playing is also added into the mixture. The overall SNR of the mixture with respect to the speech recording varies between -3 dB and -30 dB. We resampled mixtures to 16 kHz.

\subsection{Networks}
\label{imd}
We adopt the TCN module in Conv-TasNet\cite{luo2019conv} with STFT as the encoder, iSTFT as the decoder, and use 4 repeated stacks each having 6 1-D convolutional blocks in the masking network. STFTs are computed with a window size of 32 ms and a hop size of 8 ms. The effective receptive field of the model is approximately 4 s. For computational efficiency, only 8 of the channels (4 from each array) were used to train the MCSDN.

\subsection{Evaluation}
\label{ev}
We performed two subjective evaluations: one to measure the listening improvement, and the second to evaluate the importance of cACGMM in the training. We recruited 60 self-reported normal-hearing subjects who are native English speakers to participate in a listening test on Amazon Mechanical Turk \cite{paolacci2010running}. Subjects were instructed to wear headphones or earphones during the test.

The test is a simplified version of multiple stimuli with hidden reference and anchor (MUSHRA) \cite{series2014method}. We use 10 sets of male speaker samples and 10 sets of female speaker samples for the test. When evaluating each set of sounds, the subjects are instructed to listen to a 9-second unprocessed noisy speech sample first, and then listen to and rate the processed speech without knowing which algorithm had been applied. The processed speech came from 1) MCSDN, 2) MCSDN-Beamforming (mask-based MVDR beamforming), 3) MCSDN-Beamforming-MCSDN, 4) the re-mixed one with 20\% from the beamforming and 80\% from the second MCSDN, and 5) the target speech signal from cACGMM as shown in Equation~\ref{eq:cacgmm}. The subjects rate each processed speech sample on a scale with the following labels: bad (1), poor (2), fair (3), good (4), and excellent (5) on the following four aspects:
\begin{enumerate}[label=(\alph*)]
    \item \textit{Intelligibility}: How well can you recognize what the speaker is saying?
    \item \textit{Noise reduction level}: How much of the noise is removed compared to the unprocessed speech?
    \item \textit{Free of distortion}: How distortionless is the speech signal?
    \item \textit{Listening improvement}: How much the processed signal improves listening compared to the unprocessed one, e.g., how much would you like to use such a device to help them improve listening?
\end{enumerate}
Similar to MUSHRA, we used the speech signals from cACGMM as a hidden reference that were used to disqualify subjects who gave low-intelligibility and noise-reduction scores.  Because the hidden reference signals were processed from individual recordings, they contained almost no noise and should have good intelligibility scores. Then, the ratings for a set of signals from a subject were disqualified and dropped if the sum of intelligibility and noise reduction scores for the hidden reference is lower than the bottom 15\% of this summation from all subjects.

The setup for the second (cACGMM) experiment was similar to the first experiment. We provided two re-mix models with the MCSDNs trained with and without cACGMM, respectively, and asked the subjects to rate them for noise reduction and listening improvement. 


\begin{figure}[!b]
    \centering
    \includegraphics[width=1.0\columnwidth]{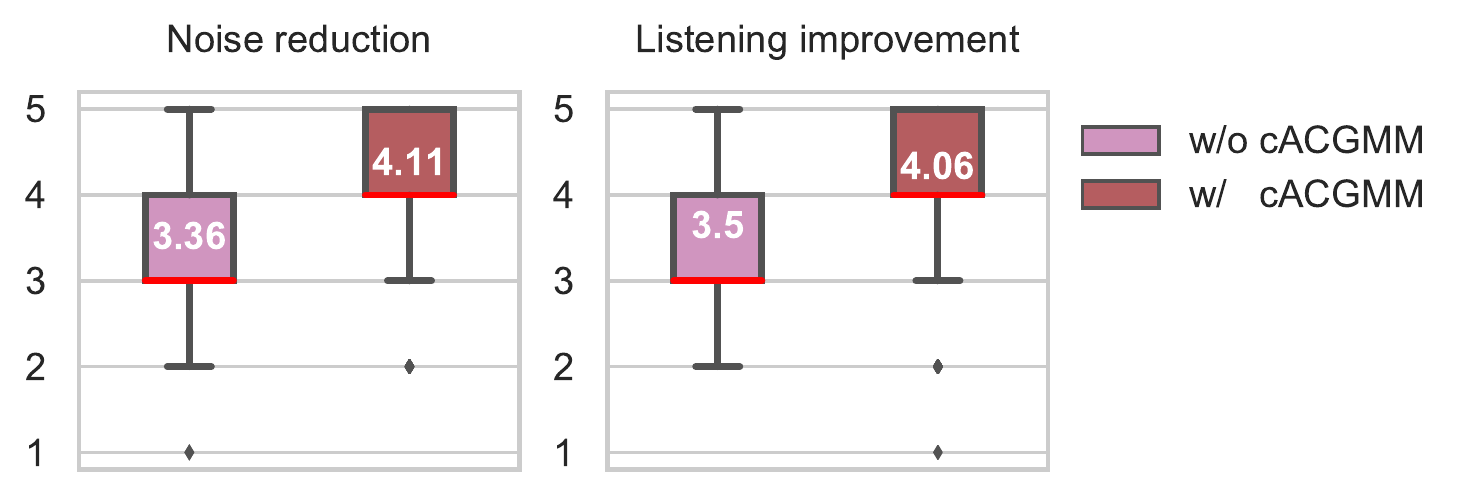}
    \caption{Comparison between the re-mix model using DNNs trained with cACGMM and without cACGMM.}
    \label{fig:r5}
\end{figure}

\section{Results and discussions}
\label{sec:result}

Figure~\ref{fig:r4} compares different models in terms of the factors described above. First, all models improve the average intelligibility score over the unprocessed mixture. We notice some subjects gave high intelligibility scores to some of the unprocessed signals even when they had low SNRs. We think this is because humans can attend to a source in the presence of multiple distracting stimuli thanks to the cocktail party effect \cite{cherry1953some}, thus they may focus on and exert themselves to understand the target speech. But overall, the strong background noise makes the speech much less intelligible. While the MCSDN has a high score for noise reduction due to the power of nonlinear models, it can cause speech distortion. If the mixture is too noisy, the model may also filter out speech components when it removes the noise. Therefore, the MCSDN only provides a slight intelligibility improvement and poor listening improvement. 

The MVDR beamformer uses linear filters to avoid distortion, which sacrifices the ability to cancel some noise. So, it has a lower noise-reduction score but a higher ``free of distortion" score. The intelligibility and listening improvement are better than those for MCSDN.

The second MCSDN, following the beamformer, reduces the residual noise noticeably but still lifts speech distortion slightly. It does not affect intelligibility and achieves slightly better listening improvement than beamforming. All metrics at the output of the second MCSDN are significantly better than the first MCSDN. 

When 20\% of the beamforming output and 80\% of the second MCSDN output are mixed as a new signal, we see it improves the intelligibility score over other models and achieves the best listening improvement. Some subjects mentioned they felt comfortable when the sound contained a little background noise. One explanation is that it is more realistic than over-denoised sound. Moreover, the beamforming output can mask the distorted components introduced by the neural networks.

The cACGMM target output achieves the highest mean scores in all aspects. This is expected because it is processed from moderately noisy recordings while the models' outputs are processed from much noisier mixtures. Here, cACGMM is shown to be able to produce good quality signals to serve as the training target for MCSDN. More than 75\% of the intelligibility, noise reduction, and overall listening improvement scores have a rating of at least 4. We notice the score for ``free of distortion" is lower, perhaps because cACGMM estimates a probabilistic time-frequency mask through spatial clustering, which may introduce nonlinear distortion. 

Figure~\ref{fig:r5} compares the output of the full re-mix model when the DNNs are trained with and without the cACGMM. We see the model trained using the cleaner target speech provided by cACGMM results in better noise reduction and thus better listening improvement. This demonstrates that cACGMM can help improve the quality of training data coming from real recordings. 

\section{Conclusion}
\label{sec:conclusion}
This paper investigates a speech denoising system using real recording data to mitigate the data mismatch problem. For training, we mixed individual recordings with reverberation and moderate noise into a mixture with multiple distractors. Instead of using speech recordings as the learning target, we applied cACGMM on individual recordings to extract clean speech signals to serve as the target for learning, which significantly improve the training dataset. For inference, we propose an MCSDN-Beamforming-MCSDN framework to take advantage of multiple microphones and balance the speech distortion and noise reduction. Subjective evaluation experiments show that this speech denoising system reduces background noise, improves speech intelligibility, and thus improves human listening. Future work includes alleviating the performance degradation from strong reverberation and tailoring the model to fit into hearing aids with low power resources.


\newpage
\bibliographystyle{IEEEtran}
{\footnotesize \bibliography{refs}}



\end{document}